\documentclass{PoS}
\usepackage{amsmath}
\title{New suppressed decays of $B^0_s$ mesons at CDF}

\ShortTitle{New suppressed decays of $B^0_s$ mesons at CDF}

\author{\speaker{Olga Norniella}\thanks{On behalf of the CDF Collaboration}\\
        Illinois University at Urbana-Champaign\\
        E-mail: \email{norniell@fnal.gov}}

\abstract{The observation of new suppressed $B^0_s$ decays, $B^0_s \rightarrow J\!/\!\psi K^*(892)^0$ 
and $B^0_s \rightarrow J\!/\!\psi K^0_S$, and 
the measurement of their branching ratios is presented. This measurement is based on an integrated luminosity of 5.9~fb$^{-1}$ of 
CDF data collected by a dedicated di-muon trigger. A cut based optimization is carried out for the observation 
of $B^0_s \rightarrow J\!/\!\psi {K^*}^0$, while a neural network is used for 
the $B^0_s \rightarrow J\!/\!\psi K^0_S$. In addition to the observation of the new decay modes,
the ratios of branching fractions to the reference $B^0$ decays are measured.
}

\FullConference{35th International Conference of High Energy Physics - ICHEP2010,\\
		July 22-28, 2010\\
		Paris France}

\begin{document}

\section*{}
While $B^0$ decays have been extensively studied at the B 
factories experiments, much less is known about $B^0_s$ decays. 
This contribution presents studies of two specific $B^0_s$ decays~\cite{PN}: $B^0_s \rightarrow J\!/\!\psi K^0_S$ and 
$B^0_s \rightarrow J\!/\!\psi  K^* (892)^0$, with $J\!/\!\psi \rightarrow \mu^+ \mu^-$, $K^0_S \rightarrow \pi^+ \pi^-$ and 
$K^*(892)^0~\rightarrow~K~\pi$.
$B^0_s \rightarrow J\!/\!\psi K^0_S$ is a CP eigenstate and has never been observed. Measurement of its lifetime directly 
probes the lifetime of the heavy mass eigenstate, $\tau_{B_{s,H}}$.  
Additionally, large samples of $B^0_s \rightarrow J\!/\!\psi K^0_S$ can be used to extract the angle $\gamma$ of the 
unitary triangle~\cite{fleischer}. 
The $B^0_s \rightarrow J\!/\!\psi {K^*}^0$ decay is yet another unobserved mode which contains an admixture of CP final states. 
With a larger data sample, an angular analysis of $B^0_s \rightarrow J\!/\!\psi {K^*}^0$ can be carried out to help 
disentangle penguin contributions in $B^0_s \rightarrow J\!/\!\psi \phi$~\cite{fleischer2}. 
In addition to the first observation of these two decays, the ratios of branching 
ratios of $B^0_s \rightarrow J\!/\!\psi K^0_S$ and $B^0_s \rightarrow J\!/\!\psi {K^*}^0$ to the reference $B^0$ decays are measured 
using the relation
\begin{equation*}
 \mathcal{B}(B^0_s \rightarrow J\!/\!\psi K)/\mathcal{B}(B^0 \rightarrow J\!/\!\psi K)=A_{rel}\times f_d/f_s\times N(B^0_s\rightarrow J\!/\!\psi K)/N(B^0\rightarrow J\!/\!\psi K), 
\vspace{-0.1cm}
\end{equation*}
\noindent
where $K$ represents $K^0_S$ or ${K^*}^0$. By measuring the ratio of the number of decays, 
$N(B^0_s \rightarrow J\!/\!\psi K)$ and $N(B^0 \rightarrow J\!/\!\psi K)$, 
from data and the relative acceptance, $A_{rel}$, between the $B^0$ and $B^0_s$ from Monte Carlo simulation (MC), the value 
$\mathcal{B}(B^0_s\rightarrow~J\!/\!\psi~K)/\mathcal{B}(B^0\rightarrow~J\!/\!\psi~K)$ is extracted by inputting the ratio of fragmentation fractions $f_s/f_d$.
 
The data used in these analyses are selected from a sample enriched in $J\!/\!\psi \rightarrow \mu^+ \mu^- $ decays, 
collected by the CDF Run II detector~\cite{cdf}. The integrated luminosity of this sample is 5.9~fb$^{-1}$. The $J\!/\!\psi$ 
dataset contains events with at least one reconstructed $J\!/\!\psi$ selected by dedicated di-muon triggers. 
In addition to the selected $J\!/\!\psi$, two tracks are combined with them in a kinematic fit to reconstruct $B^0 \rightarrow J\!/\!\psi K^0_S$ 
and $B^0 \rightarrow~J\!/\!\psi~{K^*}^0$ candidates. For the $B^0 \rightarrow J\!/\!\psi K^0_S$ analysis, the two tracks are reconstructed 
as pions and combined to define a $K^0_S$ candidate. The ${K^*}^0$ candidate for the 
$B^0 \rightarrow J\!/\!\psi {K^*}^0$ decay is reconstructed from the combination of a $\pi$ and a $K$. 

\begin{figure}[h]
\center	
  \includegraphics[angle=-90,width=.31\textwidth]{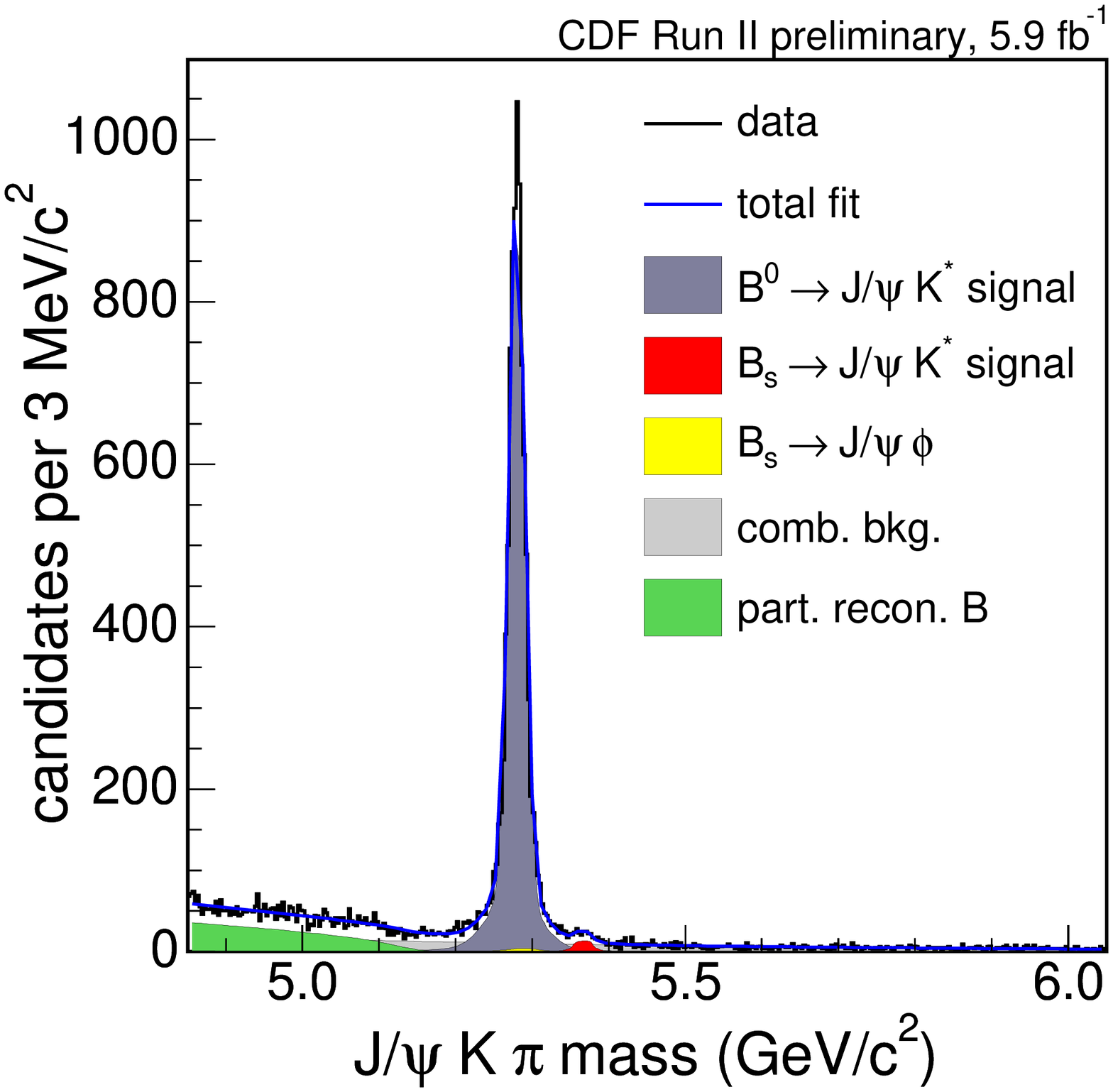}
   \includegraphics[angle=-90,width=.31\textwidth]{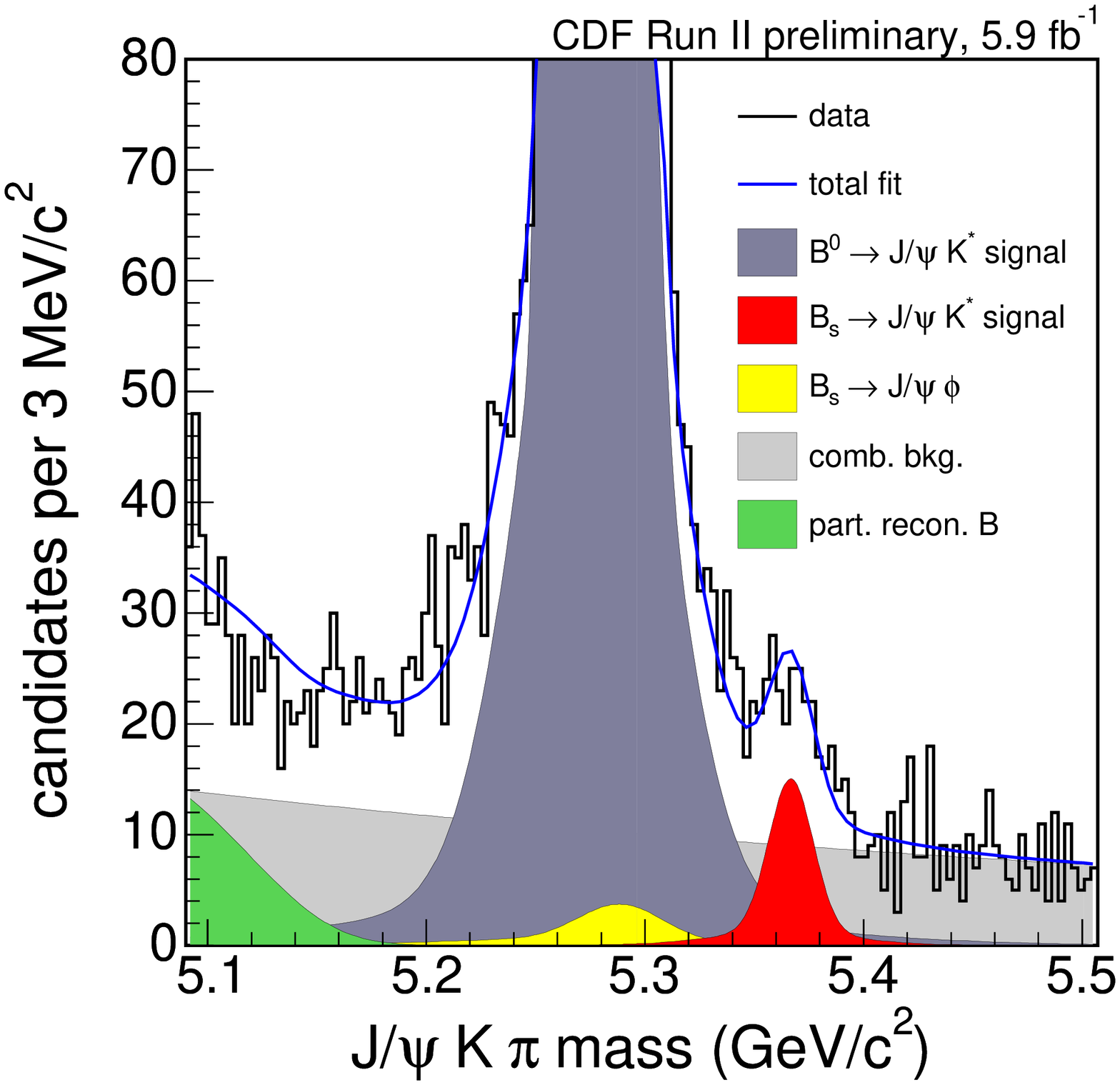}	
 \caption{\label{fig2}Invariant mass distribution for $J\!/\!\psi {K^*}^0$ and fit including the different contributions (left). 
                      The distribution is enlarged in the signal region for more detail (right).}
\end{figure}

The final event selection in the $B^0 \rightarrow J\!/\!\psi {K^*}^0$ analysis is optimized by maximizing $S/(1.5+~\sqrt B)$. 
This quantity is well suited for signal discovery as described in~\cite{punzi}. 
A simultaneous four-dimensional optimization is carried out over 4 quantities: 
$p_T(\pi)$, $p_T(K)$, transverse decay length $L_{xy}(B^0_s)$ and $B^0_s$ vertex fit probability.
For the purpose of extracting the yields of $B^0 \rightarrow J\!/\!\psi {K^*}^0$ and $B^0_s \rightarrow J\!/\!\psi {K^*}^0$ signals in the invariant mass distribution, an accurate modeling of signals and backgrounds is needed prior to the fit. 
The signal contributions are modeled with three Gaussians template obtained from a fit to $B^0$ MC. 
The $B^0_s$ template used in the final fit is 
identical to $B^0$ template, except for a shift of 86.8 MeV/$c^2$ in the mean value of the three Gaussians. 
This value corresponds to the known value~\cite{pdg} for the mass difference between $B^0_s$ and $B^0$, $\Delta (m_{B^0_s}-m_{B^0})$.
The backgrounds considered in this analysis are combinatorial background, partially reconstructed b hadron decays 
and $B^0_s \rightarrow J\!/\!\psi \phi$ decay. The first one is resulting from different sources, for example 
a real $J\!/\!\psi$ plus two random tracks, where the $J\!/\!\psi$ could be a prompt $J\!/\!\psi$ or coming from a B decay. 
Other sources that could contribute to it are fake $J\!/\!\psi$ reconstructed with prompt fake muons or 
fake muons coming from heavy flavor. 
The combinatorial background is modeled in the final fit with an exponential function.
The partially reconstructed background, which is fitted with an ARGUS function~\cite{argus},  is partially reconstructed 
b hadrons where a five-body decay occurs where a $\pi$, $K$, or $\gamma$  is not reconstructed. 
Finally, to model the $B^0_s \rightarrow J\!/\!\psi \phi$ background, a template consisting of two Gaussians, extracted from simulation, is used. 
The size of $B^0_s \rightarrow J\!/\!\psi \phi$ contribution is constrained using data.  
A binned log likelihood fit is performed to the invariant mass distributions using the templates for signals and the functions described above. Figure~\ref{fig2} shows the fit, including the different contributions. 
From the fit, the yields of the $B^0 \rightarrow J\!/\!\psi {K^*}^0$ and $B^0_s \rightarrow J\!/\!\psi {K^*}^0$ signal are 9530 $\pm$ 110 
and 151 $\pm$ 25, respectively. The statistical significance of the $B^0_s \rightarrow J\!/\!\psi {K^*}^0$ signal is 8.0$\sigma$. The 
measured ratio of yields, $N(B^0_s \rightarrow J\!/\!\psi {K^*}^0)/N(B^0 \rightarrow J\!/\!\psi {K^*}^0)$, is 
0.0159 $\pm$ 0.0022 (stat.)$\pm$0.0050 (sys.).
The systematic uncertainty is dominated by the combinatorial background contribution uncertainty, with a relative uncertainty 
for the ratio of 31.4\%. The other sources of systematic uncertainty are the signal modeling (4.4\%), uncertainty 
on $\Delta (m_{B^0_s}-m_{B^0})$ (0.1\%), combinatorial background modeling (1.3\%) and $B^0_s \rightarrow J\!/\!\psi \phi$ contribution (1.3\%). 

\begin{figure}[h]
\center	
  \includegraphics[angle=-90,width=.31\textwidth]{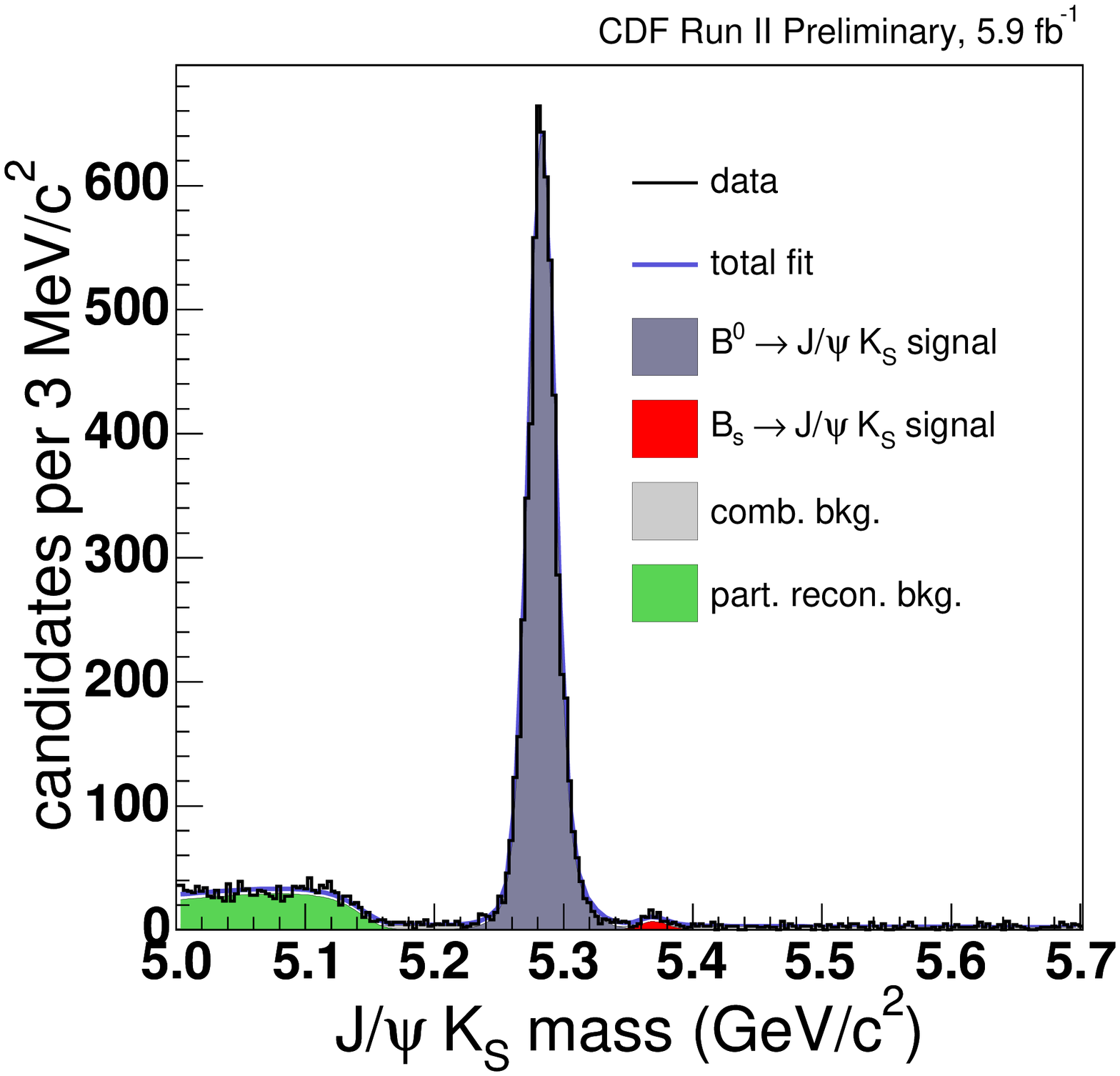}
   \includegraphics[angle=-90,width=.31\textwidth]{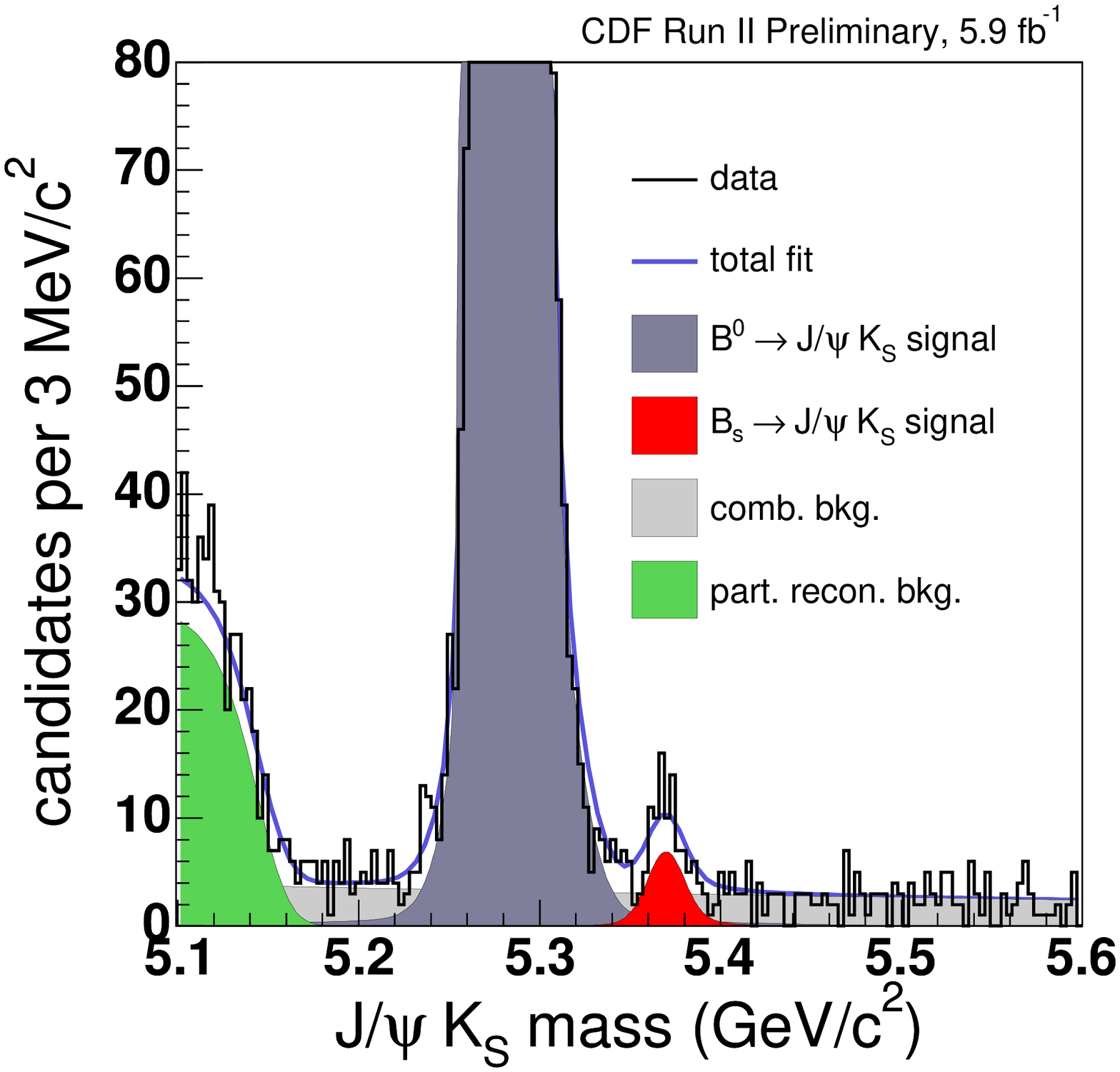}	
 \caption{\label{fig1}Invariant mass distribution for $J\!/\!\psi K^0_S$ and fit including the different contributions (left). 
                      The distribution is enlarged in the signal region for more detail (right).}
\end{figure}

The $B^0 \rightarrow J\!/\!\psi K^0_S$ decay has two main differences with respect to the $B^0 \rightarrow J\!/\!\psi {K^*}^0$ decay. First, it contains a 
$K^0_S$, which has a long lifetime. The displacement between the $K^0_S$ candidate vertex and the B candidate vertex is used in the event 
selection to reduce backgrounds from prompt sources. Moreover, the $K^0_S$ is narrow resonance, so a mass constraint can be applied and improve 
mass resolution. Second, the $B^0_s \rightarrow J\!/\!\psi K^0_S$ signal contribution is expected to be 
smaller than the $B^0_s \rightarrow J\!/\!\psi {K^*}^0$ signal contribution. Therefore it is crucial to effectively suppressing combinatorial. 
A neural network (NN) is used a discriminator. In order to train the NN, simulated $B^0_s$ events are used as signal. 
Data from the upper side band in the $B^0$ invariant mass distribution, well separated from the signal region, are used as a background data sample.
Twenty-two different kinematic and topological variables are chosen as inputs for the NN training. The selection is optimized by maximizing $S/(1.5+\sqrt B)$ and a 
likelihood fit is performed to the invariant mass distribution 
to extract the yields of $B^0 \rightarrow J\!/\!\psi K^0_S$ and 
$B^0_s \rightarrow J\!/\!\psi K^0_S$ signals. Signals and the two backgrounds are modeled with the same functional form as in 
the $B^0 \rightarrow J\!/\!\psi {K^*}^0$ analysis. 
Moreover, in the $B^0_s \rightarrow J\!/\!\psi K^0_S$ analysis , $\Lambda_b \rightarrow J\!/\!\psi \Lambda$, where $\Lambda \rightarrow p \pi$, is a 
background when the $p$ is reconstructed as a $\pi$. This contribution is suppressed using a cut in the angular distribution between $K^0_S$ and the $\pi$ with lower $p_T$. From the fit, shown in Figure~\ref{fig1}, the yields of the $B^0 \rightarrow J\!/\!\psi K^0_S$ and 
$B^0_s \rightarrow J\!/\!\psi K^0_S$ signal are determined to be 5954 $\pm$ 79 and 64 $\pm$ 14, respectively.
The statistical significance of the $B^0_s \rightarrow J\!/\!\psi K^0_S$ signal is 7.2$\sigma$. The value of $N(B^0_s \rightarrow J\!/\!\psi K^0_S)/N(B^0 \rightarrow J\!/\!\psi K^0_S)$ is 0.0108 $\pm$ 0.0019 (stat.) $\pm$ 0.0010 (sys.). The sources of the systematic uncertainty are 
similar to the other analysis. In this case the relative uncertainties for the ratio are 5.6\% from the combinatorial background 
contribution, 5.6\% from the combinatorial background modeling, 4.6\% from the signal modeling and 0.1\% from the $\Delta (m_{B^0_s}-m_{B^0})$. 

To determine the $\mathcal{B}(B^0_s \rightarrow J\!/\!\psi K)/\mathcal{B}(B^0 \rightarrow J\!/\!\psi K)$, 
where $K$ represents $K^0_S$ or ${K^*}^0$, the relative 
acceptances of $B^0 \rightarrow J\!/\!\psi K^0_S$ to $B^0_s \rightarrow J\!/\!\psi K^0_S$ and 
$B^0 \rightarrow J\!/\!\psi {K^*}^0$ to $B^0_s \rightarrow J\!/\!\psi {K^*}^0$ need to be determined from MC. 
The value for $A_{rel}$ is determined to be $A_{rel}$ = 1.012 $\pm$ 0.010 (stat.) $\pm$ 0.042 (sys.) 
for the $K^0_S$ analysis and $A_{rel}$ = 1.057 $\pm$ 0.010 (stat.) $\pm$ 0.263 (sys.) for the $K^*$ analysis.
The $B^0_s$ and $B^0$ lifetimes, B hadron $p_T$ spectrum and polarization, this last one only for the $B^0_s \rightarrow J\!/\!\psi {K^*}^0$ 
analysis, induce uncertainties in $A_{rel}$.

To determine $f_s/f_d$ , the most recent CDF measurement~\cite{fracCDF} of $f_s/(f_u+f_d)\times \mathcal{B}(D_s \rightarrow \phi \pi )$ 
is combined with the actual known value~\cite{pdg} for $\mathcal{B}(D_s \rightarrow \phi \pi )$. 
With the input of $f_s/f_d$ = 0.269 $\pm$ 0.033, the ratio of branching fractions to  the reference $B^0$ decays are:
\vspace{-0.2cm}
\begin{equation*}
 \mathcal{B}(B^0_s \rightarrow J\!/\!\psi {K^*}^0)/\mathcal{B}(B^0 \rightarrow J\!/\!\psi {K^*}^0)= 0.062 \pm 0.009 (stat.) \pm 0.025 (sys.) \pm 0.008 (frag.) ,
\end{equation*}
\vspace{-0.8cm}
\begin{equation*}
 \mathcal{B}(B^0_s \rightarrow J\!/\!\psi K^0_S)/\mathcal{B}(B^0 \rightarrow J\!/\!\psi K^0_S)= 0.041 \pm 0.007 (stat.) \pm 0.004 (sys.) \pm 0.005 (frag.). 
\end{equation*}
%\vspace{-0.5cm}
%\vspace*{0.4cm}
The world-average values for $\mathcal{B}(B^0 \rightarrow J\!/\!\psi {K^*}^0)$ and $\mathcal{B}(B^0 \rightarrow J\!/\!\psi K^0)$ are used for normalization to 
calculate the absolute branching fractions:
\vspace{-0.2cm}
\begin{equation*}
 \mathcal{B}(B^0_s \rightarrow J\!/\!\psi {K^*}^0)= (8.3 \pm 1.2 (stat.) \pm 3.3 (sys.) \pm 1.0 (frag.) \pm 0.4 (Norm.))\times10^{-5}, 
\end{equation*}
\vspace{-0.8cm}
\begin{equation*}
 \mathcal{B}(B^0_s \rightarrow J\!/\!\psi K^0)= (3.5\pm 0.6 (stat.) \pm 0.4 (sys.) \pm 0.4 (frag.) \pm 0.1 (Norm.))\times10^{-5}.
\end{equation*}

This measurement is yet another CDF contribution to the exploration of bottom-strange mesons. Further are expected with the 
10 fb$^{-1}$ data sample expected in a year from now.

\end{document}